\documentclass[aps,prd,amsmath,amssymb,reprint,superscriptaddress,nofootinbib]{revtex4-2}%
\usepackage{graphicx}
\usepackage[nopatch=footnote]{microtype}

\begin{document}
\title{The ${^{3}\!}P_{0}$ model reloaded}
\author{R. Bruschini}
\email{bruschini.1@osu.edu}
\affiliation{Department of Physics, The Ohio State University, Columbus, Ohio 43210, USA}
\author{P. Gonz\'{a}lez}
\email{pedro.gonzalez@uv.es}
\affiliation{Departamento de F\'{\i}sica Te\'{o}rica, Universidad de Valencia (UV) e IFIC
(UV - CSIC), Spain}
\author{T. Tarutina}
\email{tarutina@fisica.unlp.edu.ar}
\affiliation{IFLP, CONICET, Diagonal 113 e/ 64 y 64, 1900 La Plata, Argentina}

\begin{abstract}
We revisit the phenomenological ${^{3}\!}P_{0}$ model for the decay of quarkonium
$\left(  Q\bar{Q}\right)  $ into two open flavor mesons ($\bar{\mathfrak{M}}\mathfrak{M}%
$). We take the heavy-quark limit and derive a transition
rate between $Q\bar{Q}$ and $\bar{\mathfrak{M}}\mathfrak{M}$ to be
compared with the one calculated in studies of string breaking using lattice QCD. This comparison allows to fit the creation amplitude of a light quark-antiquark pair in the
${^{3}\!}P_{0}$ model to the string-breaking transition rate in QCD.

\end{abstract}
\maketitle

\section{Introduction}

If kinematically allowed, the dominant strong decays for quarkonia, $Q\bar{Q}$ $\left(Q=b,c\right),  $ are into open flavor two-meson states.
The dominant decay mechanism is assumed to be through the production of a
light quark-antiquark pair, $q\bar{q}$ $\left(  q=u,d,s\right)  $, and
its combination with the initial $Q\bar{Q}$ to form the open flavor
mesons $\bar{\mathfrak{M}}(Q\bar{q})$ and $\mathfrak{M}(\bar{Q}q)$. A complete description of this low-energy QCD mechanism from first principles is very difficult to achieve. Instead, simple decay models based on some ansatz for the $q\bar{q}$
production amplitude have been proposed. The most popular of them is the so called ${^{3}\!}P_{0}$ decay model, which assumes the emission out of the
vacuum of a flavor and color singlet $q\bar{q}$ pair with a spatially
constant pair poduction amplitude. The ${^{3}\!}P_{0}$ model owes its popularity to the reasonable description of data in the heavy quark meson sector as well as in the light quark meson one; see, for instance, \cite{Mic69, LY, Ono, BGS}. The phenomenological 
success of such a simple model calls for a QCD based
explanation. For quarkonium decays, this may come from lattice QCD simulations of the transition from a
colour string configuration between two static colour sources ($Q\bar{Q}$) into a pair of static-light mesons ($\bar{\mathfrak{M}}\mathfrak{M}$). 
This colour string breaking has been shown to occur \cite{Bal05,Bul19} from a $Q$-$\bar{Q}$ distance of about $1.25$ fm, and the
transition rate between $Q\bar{Q}$ and $\bar{\mathfrak{M}}\mathfrak{M}$ has been numerically evaluated \cite{Bal05,Bic20}. Then, the
comparison of this transition rate to the one derived from the ${^{3}\!}P_{0}$ model in the heavy-quark limit ($m_Q\to\infty$) may provide a QCD based justification to the observed
success of this model. This comparison is the main objective of this article
whose contents are organized as follows. In Section~2, we recall the general
${^{3}\!}P_{0}$ expression to calculate the $Q\bar{Q}\rightarrow
\bar{\mathfrak{M}}\mathfrak{M}$ decay width and take the limit $m_Q\to\infty$ as
assumed for the colour sources in calculations of string breaking using lattice QCD. In Section~3, we detail
the general procedure to extract ${^{3}\!}P_{0}$ mixing
potentials and particularize it for some selected decays. From these mixing
potentials, we calculate in Section~4 the ${^{3}\!}P_{0}$ transition rate and
compare it to the one from lattice QCD. From this comparison the suitability of
the ${^{3}\!}P_{0}$ model for describing quarkonium decays is analyzed. As
an example, in Section~5, we study bottomonium decays. In Section~6 we discuss
the application of the ${^{3}\!}P_{0}$ model to charmonium. Finally, in Section~7
we summarize our main findings and conclusions.

\section{The ${^{3}\!}P_{0}$ model for quarkonium decay}

In the ${^{3}\!}P_{0}$ model, it is assumed that the $Q\bar{Q}\rightarrow
\bar{\mathfrak{M}}\mathfrak{M}$ decay takes place in two steps. The first
step is the emission of a $\bar{q}q$
flavor and color singlet pair with the quantum numbers of the vacuum
${^{2S_{\bar{q}q}+1}\!}\left(  L_{\bar{q}q}\right)  _{J_{\bar{q}q}%
}={^{3}\!}P_{0},$ this is $S_{\bar{q}q}=1,$ $L_{\bar{q}q}=1,$
$J_{\bar{q}q}=0.$ 
The second step is the combination of the color singlet
$\bar{q}q$ with the color singlet $Q\bar{Q}$ giving rise to the
$\bar{\mathfrak{M}}\left(  Q\bar{q}\right)  $ and $\mathfrak{M}%
\left(  \bar{Q}q\right)  $ mesons.

It is customarily assumed that the amplitude for the creation of $q\bar{q}$ is a constant $\gamma$ \cite{Mic69, LY, Ono, BGS}. In this case, the
width $\Gamma_{{^{3}\!}P_{0}}$ for the process in the rest frame of the decaying quarkonia reads \cite{Ono}%
\begin{align}
& \Gamma_{{^{3}\!}P_{0}}\left(  Q\bar{Q}\rightarrow\bar{\mathfrak{M}%
}\mathfrak{M}\right)  \label{3P0width}\\
& =2\pi\frac{E_{\mathfrak{M}}E_{\bar{\mathfrak{M}}}}{m_{Q\bar{Q}}%
}k\gamma^{2}\left\vert \left\langle I_{\bar{\mathfrak{M}}},\tau_{\bar{\mathfrak{M}}};I_{\mathfrak{M}},\tau_\mathfrak{M}\right.  \left\vert
I_{Q\bar{Q}}, \tau_{Q\bar{Q}}\right\rangle \right\vert ^{2}%
\nonumber\\
& \left[
\begin{array}
[c]{ccc}%
I_{Q} & I_{\bar{Q}} & I_{Q\bar{Q}}\\
I_{\bar{q}} & I_{q} & 0\\
I_{\bar{\mathfrak{M}}} & I_{\mathfrak{M}} & I_{Q\bar{Q}}%
\end{array}
\right]  ^{2}\left[
\begin{array}
[c]{ccc}%
S_{Q} & S_{\bar{Q}} & S_{Q\bar{Q}}\\
S_{\bar{q}} & S_{q} & 1\\
S_{\bar{\mathfrak{M}}} & S_{\mathfrak{M}} & S_{\bar{\mathfrak{M}%
}\mathfrak{M}}%
\end{array}
\right]  ^{2}\nonumber\\
& \left(
\begin{array}
[c]{c}%
\left[
\begin{array}
[c]{ccc}%
L_{Q\bar{Q}} & S_{Q\bar{Q}} & J_{Q\bar{Q}}\\
1 & 1 & 0\\
L_{Q\bar{Q}}-1 & S_{\bar{\mathfrak{M}}\mathfrak{M}} & J_{Q\bar
{Q}}%
\end{array}
\right]  ^{2}\left\vert \mathcal{J}_{-}(k)\right\vert ^{2}+\\
\left[
\begin{array}
[c]{ccc}%
L_{Q\bar{Q}} & S_{Q\bar{Q}} & J_{Q\bar{Q}}\\
1 & 1 & 0\\
L_{Q\bar{Q}}+1 & S_{\bar{\mathfrak{M}}\mathfrak{M}} & J_{Q\bar
{Q}}%
\end{array}
\right]  ^{2}\left\vert \mathcal{J}_{+}(k)\right\vert ^{2}%
\end{array}
\right),  \nonumber
\end{align}
where $E$ stands for the energy, $k$ for the relative $\mathfrak{M}$-$\bar{\mathfrak{M}}$ momentum, $m$ for the mass, $I$ for isospin, $\tau$ for
its third component, $S$ for spin, $L$ for orbital angular momentum, and $J$ for
total angular momentum. We have also introduced the shorthand%
\begin{align*}
&  \left[
\begin{array}
[c]{ccc}%
j_{1} & j_{2} & j_{3}\\
j_{4} & j_{5} & j_{6}\\
j_{7} & j_{8} & j_{9}%
\end{array}
\right]  \\
&  \equiv\sqrt{\left(  2j_{3}+1\right)  \left(  2j_{6}+1\right)  \left(
2j_{7}+1\right)  \left(  2j_{8}+1\right)  }\left\{
\begin{array}
[c]{ccc}%
j_{1} & j_{2} & j_{3}\\
j_{4} & j_{5} & j_{6}\\
j_{7} & j_{8} & j_{9}%
\end{array}
\right\},
\end{align*}
where $\left\{
\begin{smallmatrix}
j_{1} & j_{2} & j_{3}\\
j_{4} & j_{5} & j_{6}\\
j_{7} & j_{8} & j_{9}%
\end{smallmatrix}
\right\}  $ is the Wigner $9j$-symbol.

The factors $\mathcal{J}_{+}(k)$ and $\mathcal{J}_{-}(k)$ inside Eq.~\eqref{3P0width} are given by
\begin{align}
\mathcal{J}_{-}(k)&=\frac{i^{L_{Q\bar{Q}}}}{\sqrt{8}}\sqrt{\frac{3L_{Q\bar{Q}%
}}{2L_{Q\bar{Q}}-1}}\mathcal{I}_{-}(k)\label{J1}\\
\mathcal{J}_{+}(k)&=\frac{i^{L_{Q\bar{Q}}}}{\sqrt{8}}\sqrt{\frac{3\left(
L_{Q\bar{Q}}+1\right)  }{2L_{Q\bar{Q}}+3}}\mathcal{I}_{+}%
(k),\label{J2}%
\end{align}
where $\mathcal{I}_{+}(k)$ and $\mathcal{I}_{-}(k)$ are integrals involving the radial wave functions of the mesons in the initial and final states. They are expressed as
\begin{align}
& \mathcal{I}_{\pm}(k)\label{I12}\\
& =\int \mathrm{d}r\,r^{2}\int \mathrm{d}p\, p^{2}\phi_{\bar{\mathfrak{M}}%
}^{\ast}\left(  p\right)  \phi_{\mathfrak{M}}^{\ast}(p)\psi_{L_{Q\bar{Q}%
}}(r)\nonumber\\
& \left[ p j_{1}\left(  pr\right)  j_{L_{Q\bar{Q}}\pm1}\left(
\hat{h}_{Q}kr\right)  \pm \hat{h}_{q}kj_{0}(pr)j_{L_{Q\bar{Q}}}\left(  \hat{h}_{Q}%
kr\right)  \right],  \nonumber
\end{align}
where $r$ is the $Q$-$\bar{Q}$ relative distance, $\psi_{L_{Q\bar{Q}}%
}(r)$ the radial $Q\bar{Q}$ wave function in the initial state, $p$ the modulus of the
relative momentum of the quark $Q$ and antiquark $\bar{q}$ in
$\bar{\mathfrak{M}}$,%
\begin{align}
\hat{h}_{Q}&\equiv\frac{m_{Q}}{m_{Q}+m_{q}},\label{h1}\\
\hat{h}_{q}&\equiv\frac{m_{q}}{m_{Q}+m_{q}}\label{h2},
\end{align}
and $\phi_{\mathfrak{M}}(p)=\phi_{\bar
{\mathfrak{M}}}\left( p\right)$ is the radial wave function of the open-flavor mesons in momentum space.

For decays into $S$-wave mesons, as it will be the case in our analysis, $\phi_{\mathfrak{M}%
}(p)$ is given by
\begin{equation}
\phi_{\mathfrak{M}}(p)=\sqrt{\frac{2}{\pi}}\int_{0}^{\infty}\mathrm{d}r_{\bar{Q}%
q}\,r_{\bar{Q}q}^{2}\psi_{\mathfrak{M}}\left(  r_{\bar{Q}%
q}\right)  j_{0}\left( p r_{\bar{Q}q}\right),  \label{um}%
\end{equation}
where $\psi_{\mathfrak{M}}\left(  r_{\bar{Q}q}\right)  $ is the radial
wave function of the open-flavor meson and $j_{n}\left(  z\right)  $ the spherical Bessel function.

\bigskip

For $m_{Q}\gg m_{q}$ we can take the heavy-quark limit $m_Q\to\infty$. We will refer to the ${^3\!}P_0$ model in such a limit as the \emph{static ${^3\!}P_0$ model}. Then, one has
\begin{equation}
\hat{h}_{Q}  \simeq1, \qquad \hat{h}_{q}  \simeq0, \label{Aph}
\end{equation}
so that
\begin{align}
& \mathcal{I}_{\pm}^\textup{st}(k)\label{I12red}\\
& =\int \mathrm{d}r\,r^{2}\int \mathrm{d}p\,p^{2}\phi_{\bar{\mathfrak{M}}%
}^{\ast}\left(  p\right)  \phi_{\mathfrak{M}}^{\ast}(p)\psi_{L_{Q\bar{Q}%
}}(r)\nonumber\\
& pj_{1}\left(  pr\right)  j_{L_{Q\bar{Q}}\pm1}\left(  kr\right)
\nonumber\\
& \equiv\int \mathrm{d}r\,r^{2}\psi_{L_{Q\bar{Q}}}(r)a\left(  r\right)
j_{L_{Q\bar{Q}}\pm1}\left(  kr\right), \nonumber
\end{align}
where the superscript ``st'' stands for static and we have defined%
\begin{equation}
a\left(  r\right)  \equiv\int \mathrm{d}p\,p^{3}\phi_{\bar{\mathfrak{M}}%
}^{\ast}\left(  p\right)  \phi_{\mathfrak{M}}^{\ast}(p)j_{1}\left(  pr\right).
\label{aint}%
\end{equation}

\bigskip

Consistently with the existing literature \cite{Mic69, LY, Ono, BGS}, we assume for the open flavor mesons a $1S$ harmonic oscillator
radial wave function. Following the notation in \cite{LY} we write%
\begin{equation}
\psi_{\mathfrak{M}}\left(  r_{\bar{Q}q}\right)  =\frac{2}{\pi^{\frac
{1}{4}}R^{\frac{3}{2}}}e^{-\frac{r_{\bar{Q}q}^{2}}{2R^{2}}}, \label{howf}%
\end{equation}
where $R$ is a dimensional parameter to be fitted from data (notice that the
root mean square radius (rmsr) of the open flavor meson is $\left\langle
r_{\bar{Q}q}^{2}\right\rangle ^{\frac{1}{2}}=\sqrt{\frac{3}{2}}R$).

Then%
\begin{equation}
\phi_{\mathfrak{M}}(p)=\frac{2R^{\frac{3}{2}}}{\pi^{\frac{1}{4}}}%
e^{-\frac{p^{2}R^{2}}{2}} \label{howfmom}%
\end{equation}
and using%
\begin{equation}
j_{1}(z)=\frac{\sin z}{z^{2}}-\frac{\cos z}{z}, \label{bes}%
\end{equation}
we obtain%
\begin{equation}
a\left(  r\right)  =\frac{1}{2R^{2}}re^{-\frac{r^{2}}{4R^{2}}}. \label{aho}%
\end{equation}

If we now group the angular momentum dependence through the coefficients%
\begin{align}
C_{-}  & \equiv\left[
\begin{array}
[c]{ccc}%
\frac{1}{2} & \frac{1}{2} & S_{Q\bar{Q}}\\
\frac{1}{2} & \frac{1}{2} & 1\\
S_{\bar{\mathfrak{M}}} & S_{\mathfrak{M}} & S_{\bar{\mathfrak{M}%
}\mathfrak{M}}%
\end{array}
\right] \left[
\begin{array}
[c]{ccc}%
L_{Q\bar{Q}} & S_{Q\bar{Q}} & J_{Q\bar{Q}}\\
1 & 1 & 0\\
L_{Q\bar{Q}}-1 & S_{\bar{\mathfrak{M}}\mathfrak{M}} & J_{Q\bar
{Q}}%
\end{array}
\right] \label{coef-}\\
&\sqrt{ \frac{3L_{Q\bar{Q}}}{2L_{Q\bar{Q}}-1}}\nonumber
\end{align}%
\begin{align}
C_{+}  & \equiv\left[
\begin{array}
[c]{ccc}%
\frac{1}{2} & \frac{1}{2} & S_{Q\bar{Q}}\\
\frac{1}{2} & \frac{1}{2} & 1\\
S_{\bar{\mathfrak{M}}} & S_{\mathfrak{M}} & S_{\bar{\mathfrak{M}%
}\mathfrak{M}}%
\end{array}
\right]\left[
\begin{array}
[c]{ccc}%
L_{Q\bar{Q}} & S_{Q\bar{Q}} & J_{Q\bar{Q}}\\
1 & 1 & 0\\
L_{Q\bar{Q}}+1 & S_{\bar{\mathfrak{M}}\mathfrak{M}} & J_{Q\bar
{Q}}%
\end{array}
\right] \label{coef+}\\
& \sqrt{\frac{3\left(  L_{Q\bar{Q}}+1\right)  }{2L_{Q\bar{Q}}+3}}\nonumber
\end{align}
we can write the decay width as%
\begin{align}
& \Gamma_{{^{3}\!}P_{0}}^\textup{st}\left(  Q\bar{Q}\rightarrow\bar
{\mathfrak{M}}\mathfrak{M}\right)  \label{width1}\\
& =\frac{\pi}{4}\frac{E_{\mathfrak{M}}E_{\bar{\mathfrak{M}}}k}{m_{Q\bar{Q}}%
}\gamma^{2}\left\vert \left\langle I_{\bar{\mathfrak{M}}}, \tau_{\bar
{\mathfrak{M}}}; I_{\mathfrak{M}}, \tau_{\mathfrak{M}}\right.  \left\vert
I_{Q\bar{Q}}, \tau_{Q\bar{Q}}\right\rangle \right\vert ^{2}%
\nonumber\\
& \left[
\begin{array}
[c]{ccc}%
I_{Q} & I_{\bar{Q}} & I_{Q\bar{Q}}\\
I_{\bar{q}} & I_{q} & 0\\
I_{\bar{\mathfrak{M}}} & I_{\mathfrak{M}} & I_{Q\bar{Q}}%
\end{array}
\right]  ^{2}\left(  C_{-}^{2}\left\vert \mathcal{I}_{-}^\textup{st}(k)\right\vert
^{2}+C_{+}^{2}\left\vert \mathcal{I}_{+}^\textup{st}(k)\right\vert ^{2}\right).
\nonumber
\end{align}

\bigskip

We shall centre first in bottomonium
decays for which we expect the limit $m_Q\to\infty$ to be more accurate. More concretely, we
shall consider%
\begin{equation}
b\bar{b}\rightarrow B^{+}B^{-},B^{0}\bar{B}^{0}.%
\end{equation}
Then $I_{b}=0=I_{\bar{b}},$ $I_{b\bar{b}}=0,$ $q=u,d$ and
$I_{q}=\frac{1}{2}=I_{\bar{q}},$ $I_{\bar{B}}=\frac{1}{2}=I_{B}$ and $\tau_B=\pm\frac{1}{2}=-\tau_{\bar{B}}$, so
that
\begin{equation}
\left\vert \left\langle \frac{1}{2},\pm\frac{1}{2};\frac{1}{2},\mp\frac{1}{2}\right.  \left\vert 0,0\right\rangle \right\vert ^{2}=\frac{1}{2} \label{is1}
\end{equation}
and
\begin{equation}
\left[
\begin{array}
[c]{ccc}%
0 & 0 & 0\\
\frac{1}{2} & \frac{1}{2} & 0\\
\frac{1}{2} & \frac{1}{2} & 0
\end{array}
\right]  ^{2}=1 \label{is2}.%
\end{equation}
As for the open-flavor meson spin $S_{\bar{B}}=0=S_{B}.$

\bigskip

Hence, for instance, the expression for the $b\bar{b}\rightarrow
B^{0}\bar{B}^{0}$ width is%
\begin{align}
&  \Gamma_{{^{3}\!}P_{0}}^\textup{st}\left(  b\bar{b}\rightarrow B^{0}\bar
{B}^{0}\right)  \label{w+-}\\
&  =\frac{\pi}{8}\frac{E_{B^{0}}E_{\bar{B}^{0}}}{m_{b\bar{b}}}k\gamma
^{2}\left(  C_{-}^{2}\left\vert \mathcal{I}_{-}^\textup{st}(k)\right\vert ^{2}%
+C_{+}^{2}\left\vert \mathcal{I}_{+}^\textup{st}(k)\right\vert ^{2}\right).  \nonumber
\end{align}
In Table I we list the nonvanishing angular momentum coefficients for some
selected decays, which we use as examples in what follows.

\begin{table}%
\begin{ruledtabular}
\begin{tabular}
[c]{ccccc}%
$\left(  J^{PC}\right)  _{Q\bar{Q}}\rightarrow\left(  J^{P}\right)
_{\bar{\mathfrak{M}}}\left(  J^{P}\right)  _{\mathfrak{M}}$ & $L_{Q\bar{Q}}$ & $S_{Q\bar
{Q}}$ & $C_{-}$ & $C_{+}$\\
\hline
$0^{++}\rightarrow0^{-}0^{-}$ & $1$ & $1$ & $\frac{1}{2}$ & $0$\\
$1^{--}\rightarrow0^{-}0^{-}$ & $0$ & $1$ & $0$ & $\frac{1}{2\sqrt{3}}$\\
$1^{--}\rightarrow0^{-}0^{-}$ & $2$ & $1$ & $\frac{1}{\sqrt{6}}$ & $0$
\end{tabular}
\end{ruledtabular}
\caption{Angular momentum coefficients for some selected decays.}
\label{I}%
\end{table}

\section{${^{3}\!}P_{0}$ mixing potentials}

The $Q\bar{Q}\rightarrow\bar{\mathfrak{M}}\mathfrak{M}$ decay width
can be alternatively calculated from a $Q\bar{Q}$-$\bar{\mathfrak{M}%
}\mathfrak{M}$ interaction, or mixing, potential \cite{Fan61,Eic78}. More
explicitly, if we consider for the sake of simplicity decays where $\bar
{\mathfrak{M}}\mathfrak{M}$ is characterized by only one $\left(
L_{\bar{\mathfrak{M}}\mathfrak{M}},S_{\bar{\mathfrak{M}}%
\mathfrak{M}}\right)  $ pair of values and $Q\bar{Q}$ by only one
$\left(  L_{Q\bar{Q}},S_{Q\bar{Q}}\right)  $ pair of values, then the
width can be written as%
\footnote{For a detailed derivation of this expression, see
\cite{Bru21a}. Notice that in this reference an educated guess for the mixing
potential, missing the angular momentum coefficients, was used.}
\begin{align}
& \Gamma\left(  Q\bar{Q}\rightarrow\bar{\mathfrak{M}}\mathfrak{M}%
\right)  \label{wmix}\\
& =4\mu_{\bar{\mathfrak{M}}\mathfrak{M}}k\left\vert \int \mathrm{d}r\,%
r^{2}j_{L_{\bar{\mathfrak{M}}\mathfrak{M}}}(kr)V^{Q\bar
{Q}\rightarrow\bar{\mathfrak{M}}\mathfrak{M}}\left(  r)\right)
\psi_{L_{Q\bar{Q}}}(r)\right\vert ^{2},\nonumber
\end{align}
where $\mu_{\bar{\mathfrak{M}}\mathfrak{M}}$ is the reduced mass of the
$\bar{\mathfrak{M}}\mathfrak{M}$ system and $V^{Q\bar{Q}%
\rightarrow\bar{\mathfrak{M}}\mathfrak{M}}$ the mixing potential.

In particular,
\begin{align}
& \Gamma\left(  b\bar{b}\rightarrow B^{0}\bar{B}^{0}\right)
\label{wmix+-}\\
& =4\mu_{B^{0}\bar{B}^{0}}k\left\vert \int \mathrm{d}r\,r^{2}j_{L_{B^{0}%
\bar{B}^{0}}}(kr)V^{b\bar{b}\rightarrow B^{0}\bar{B}^{0}%
}\left(  r\right)  \psi_{L_{b\bar{b}}}(r)\right\vert ^{2}.\nonumber
\end{align}
Then, the comparison of (\ref{w+-}) with (\ref{wmix+-}) allows for the
definition of ${^{3}\!}P_{0}$ mixing potentials. For example, for
the $1^{--}\rightarrow0^{-}0^{-}$ case with $L_{b\bar{b}}=0$ one has
$S_{b\bar{b}}=1,$ $J_{b\bar{b}}=1,$ $L_{B^{0}\bar{B}^{0}}=1$
and%
\begin{align}
& \Gamma_{{^{3}\!}P_{0}}^\textup{st}\left(  1^{--}\rightarrow0^{-}0^{-}\right)
_{L_{b\bar{b}}=0}\label{Case11}\\
& =\frac{\pi}{8}\frac{E_{B^{0}}E_{\bar{B}^{0}}}{m_{b\bar{b}}}k\gamma^{2}%
C_{+}^{2}\left\vert \mathcal{I}_{+}^\textup{st}(k)\right\vert ^{2}\nonumber\\
& =\frac{\pi}{96}\mu_{B^{0}\bar{B}^{0}}k\gamma^{2}\left(  \int dr\text{
}r^{2}j_{1}(kr)a(r)\psi_{0}(r)\right)  ^{2}\nonumber,
\end{align}
where in the last line we have substituted the value of $C_+$ from Table~\ref{I} and we have used that $\mu_{\bar{\mathfrak{M}%
}\mathfrak{M}}\simeq\frac{E_{\bar{\mathfrak{M}}}E_{\mathfrak{M}}%
}{m_{Q\bar{Q}}}$ in the $m_Q\to\infty$ limit. Comparing the above expression with%
\begin{align}
& \Gamma_{{^{3}\!}P_{0}}^\textup{st}\left(  1^{--}\rightarrow0^{-}0^{-}\right)
_{L_{b\bar{b}}=0}\label{Case12}\\
& =4\mu_{B^{0}\bar{B}^{0}}k\left(  \left\vert \int \mathrm{d}r\,r^{2}%
j_{1}(kr)V_{{^{3}\!}P_{0}}^{\left(  1^{--}\rightarrow0^{-}0^{-}\right)
_{L_{b\bar{b}}=0}}\psi_{0}(r)\right\vert ^{2}\right)  \nonumber,
\end{align}
we obtain%
\begin{equation}
\left\vert V_{_{{^{3}\!}P_{0}}}^{\left(  1^{--}\rightarrow0^{-}0^{-}\right)
_{L_{b\bar{b}}=0}}\right\vert =\frac{1}{8}\sqrt{\frac{\pi}{6}}\gamma
a(r).\label{V3P01}%
\end{equation}
By proceeding in the same manner for the selected decays one can easily obtain%
\begin{equation}
\left\vert V_{{^{3}\!}P_{0}}^{\left(  1^{--}\rightarrow0^{-}0^{-}\right)
_{L_{b\bar{b}}=2}}\right\vert =\frac{1}{8}\sqrt{\frac{\pi}{3}}\gamma
a(r)\label{V3P02}%
\end{equation}
and%
\begin{equation}
\left\vert V_{{^{3}\!}P_{0}}^{0^{++}\rightarrow0^{-}0^{-}}\right\vert =
\frac{1}{4}\sqrt{\frac{\pi}{2}}\gamma a(r).\label{V3P03}%
\end{equation}

\section{${^{3}\!}P_{0}$ transition rate}

In lattice QCD studies of string breaking the transition rate $g(r)$ between
$Q\bar{Q}$ and $\bar{\mathfrak{M}}\mathfrak{M}$ has been evaluated
from a $2\times2$ correlation matrix with static and light quarks
\cite{Bal05,Bic20}. By making use of the so called diabatic approach in QCD
\cite{Bic20,Bru20} this transition rate has been related to the $Q\bar
{Q}$-$\bar{\mathfrak{M}}\mathfrak{M}$ mixing potentials \cite{Bru23}. More
concretely, the mixing potential for any particular $Q\bar{Q}%
\rightarrow\bar{\mathfrak{M}}\mathfrak{M}$ decay can be derived from the
transition rate $g(r)$ by multiplying it by appropriate angular momentum
coefficients. Some of these coefficients have been tabulated in \cite{Bru23}, and a general formula has been derived in \cite{Braa24}. It turns out that, up to an arbitrary sign, these angular momentum
coefficients are the same $C_{-}$ and $C_{+}$
appearing in the ${^{3}\!}P_{0}$ model for the selected decays; see Table \ref{I}.

An additional technical comment is in order. As shown in
\cite{Bru21b}, for the case of degenerate $\bar{\mathfrak{M}}%
\mathfrak{M}$ states, as it is very approximately the case for $B^{+}B^{-}%
$\ and $B^{0}\bar{B}^{0},$ the transition rate $g(r)$ multiplied by the
angular momentum coefficients gives $\sqrt{2}$ times the $b\bar
{b}\rightarrow B^{+}B^{-}$ or $b\bar{b}\rightarrow B^{0}\bar{B}^{0}$
mixing potential.

\bigskip

The other way around, we can use the angular momentum coefficients to extract
from the ${^{3}\!}P_{0}$ mixing potentials the ${^{3}\!}P_{0}$
transition rate $g_{{^{3}\!}P_{0}}(r).$ Thus, from the decays previously selected,
one has%
\begin{align}
\left\vert g_{{^{3}\!}P_{0}}(r)\right\vert  &  =\left\vert \frac{\sqrt{2}%
V_{{^{3}\!}P_{0}}^{\left(  1^{--}\rightarrow0^{-}0^{-}\right)  _{L_{b\bar{b}%
}=0}}}{\frac{1}{2\sqrt{3}}}\right\vert \label{g3P0}\\
&  =\left\vert \frac{\sqrt{2}V_{{^{3}\!}P_{0}}^{\left(  1^{--}\rightarrow
0^{-}0^{-}\right)  _{L_{b\bar{b}}=2}}}{\frac{1}{\sqrt{6}}}\right\vert
\nonumber\\
&  =\left\vert \frac{\sqrt{2}\left(  V_{{^{3}\!}P_{0}}^{0^{++}\rightarrow
0^{-}0^{-}}\right)  }{\frac{1}{2}}\right\vert \nonumber\\
&  =\frac{\sqrt{\pi}}{4}\gamma a(r)\nonumber\\
&  =\gamma\frac{\sqrt{\pi}}{8}\frac{r}{R^{2}}e^{-\frac{r^{2}}{4R^{2}}}.\nonumber
\end{align}

\section{The ${^{3}\!}P_{0}$ vs the string breaking transition rate}

\begin{figure}
\centering
\includegraphics[width=\linewidth]{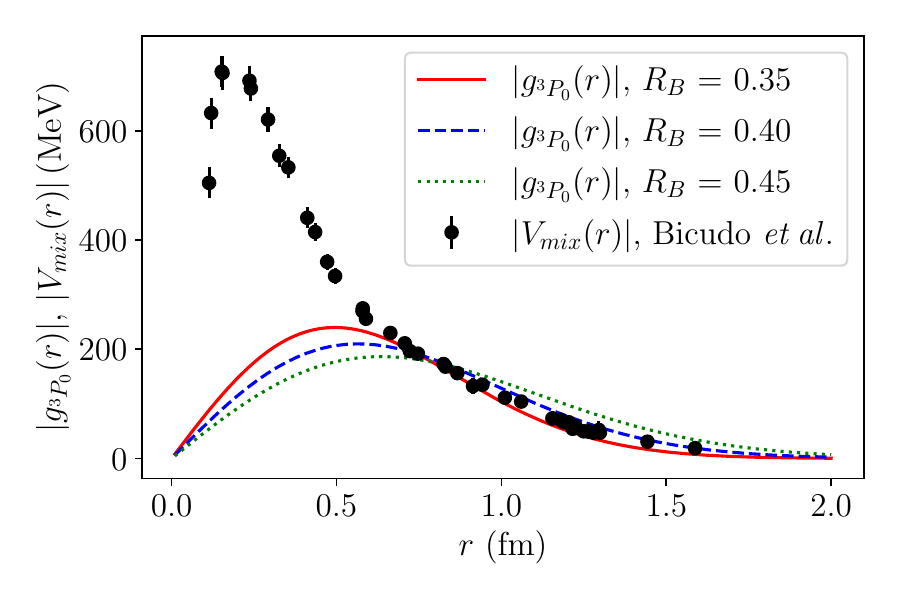}\caption{Transition rate from
the static ${^{3}\!}P_{0}$ model against lattice results from \cite{Bic20}.}%
\label{figure1}%
\end{figure}

As we expect the $Q\bar{Q}\rightarrow\bar{\mathfrak{M}}\mathfrak{M}$
decay to take dominantly place when the $Q\bar{Q}$ separation reaches the
string breaking distance, which has been estimated to be $1.25$ fm in
\cite{Bal05}, we choose $\gamma$ so that $\left\vert g_{{^{3}\!}P_{0}%
}(r)\right\vert $ fits $\left\vert g(r)\right\vert $ at the long range. The
result is shown in Fig. \ref{figure1} where we compare the calculated
$\left\vert g_{{^{3}\!}P_{0}}(r)\right\vert $ for $\gamma=2.24$ and several values
of $R_{B}$ around $0.4$ fm, in the expected range of the rmsr of the $B$ meson
from quark model descriptions, to $\left\vert g(r)\right\vert $ extracted from
the last column of Table IV in \cite{Bic20} (notice that in this reference
$g(r)$ is called $V_\textup{mix})$.

This comparison shows that for $\gamma$ close to the chosen value
$\gamma=2.24$ the ${^{3}\!}P_{0}$ decay model provides an accurate description of
the transition rate\ at the long range, i.e., for distances $r>0.6$ fm,
whereas the short range description is clearly deficient. As a matter of fact,
the fitting of the short range would require values of $R_{B}$ around $0.14$
fm, far out of the phenomenological ones, and would spoil the long range description.

\begin{figure}
\centering
\includegraphics[width=\linewidth]{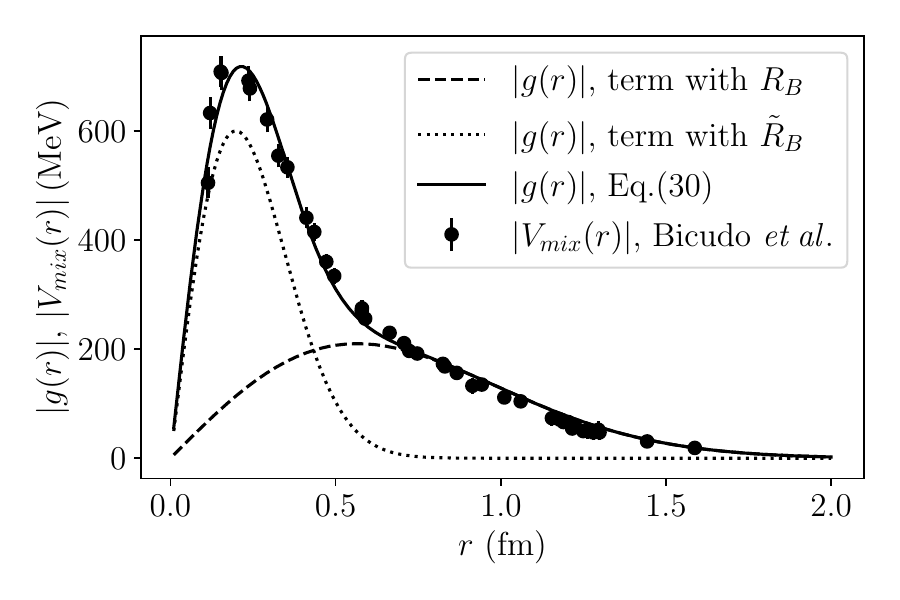}\caption{Parametrization fit of
the transition rate from \cite{Bic20}.}%
\label{figure2}%
\end{figure}

It is however possible to get a good fit of $\left\vert g(r)\right\vert $ at
both ranges by combining the results for $R_{B}\simeq0.4$ fm and
$\widetilde{R}_{B}\simeq0.14$ fm, through the parametrization%
\begin{equation}
\left\vert g(r)\right\vert \simeq\gamma\frac{\sqrt{\pi}}{8}\left(  \frac{r}{R_{B}^{2}%
}e^{-\frac{r^{2}}{4R_{B}^{2}}}+\frac{r}{\widetilde{R}_{B}^{2}}e^{-\frac{r^{2}%
}{4\widetilde{R}_{B}^{2}}}\right)  \label{parg}%
\end{equation}
with $\gamma=2.24$. This fit is shown in Fig. \ref{figure2}.

\begin{figure}
\centering
\includegraphics[width=\linewidth]{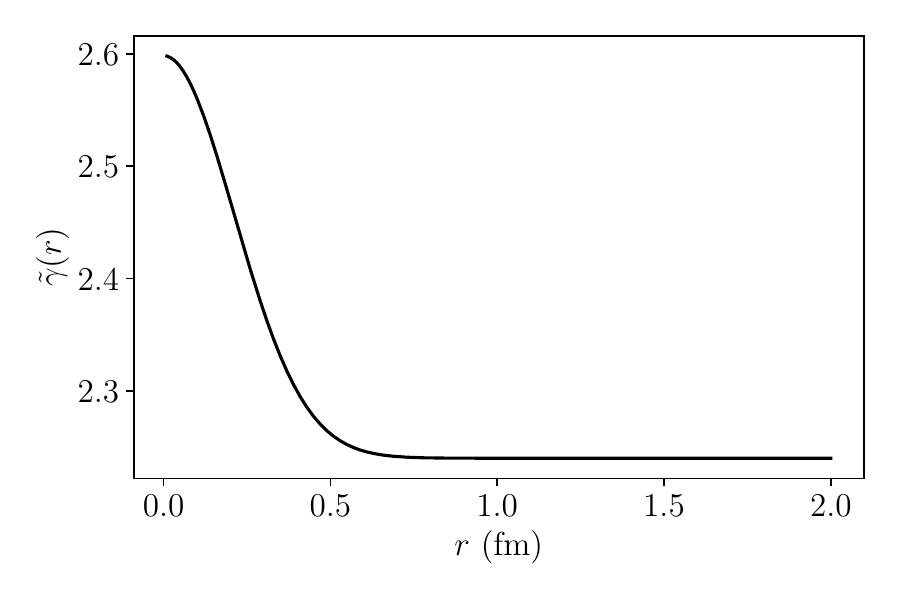}\caption{Spatial
dependent ${^{3}\!}P_{0}$ pair creation amplitude.}%
\label{figure3}%
\end{figure}

Actually, this parametrization (\ref{parg}) is quite similar to the numerical
one given in \cite{Bic20} and admits an interpretation in the framework of the
${^{3}\!}P_{0}$ model by rewriting it as%
\begin{equation}
\left\vert g(r)\right\vert \simeq\widetilde{\gamma}\left(  r\right)  \frac{\sqrt
{\pi}}{8}\frac{r}{R_{B}^{2}}e^{-\frac{r^{2}}{4R_{B}^{2}}}, \label{gammar1}%
\end{equation}
where
\begin{equation}
\widetilde{\gamma}\left(  r\right)  \equiv\gamma\left(  1+\frac{R_{B}^{2}%
}{\widetilde{R}_{B}^{2}}e^{-\frac{r^{2}}{4}\left(  \frac{1}{\widetilde{R}%
_{B}^{2}}-\frac{1}{R_{B}^{2}}\right)  }\right)  \label{gammar2}%
\end{equation}
is a $r$-dependent amplitude for the creation of the
$q\bar{q}$ pair. In Fig. \ref{figure3} we plot $\widetilde{\gamma}\left(
r\right)  $, making it clear that for $r>0.6$ fm it is almost constant.

It should be pointed out that the fitted value of $\gamma$ is
quite the same used in reference \cite{Ono} to describe charmonium
decays.

\section{Analysis of the ${^{3}\!}P_{0}$ model in bottomonium}

From the accurate ${^{3}\!}P_{0}$ description of the string breaking transition
rate between $Q\bar{Q}$ and $\bar{\mathfrak{M}}\mathfrak{M}$ in the
long range we\ may expect the suitability of the ${^{3}\!}P_{0}$ model for decays
where the dominant contribution to the width comes from $r>0.6$ fm. In order
to check this we calculate the width for the decay $\Upsilon\left(  4S\right)
\rightarrow B^{-}B^{+},B^{0}\bar{B}^{0}.$ Following the Particle Data
Book \cite{PDG} the $\Upsilon\left(  4S\right)$ is the only $b\bar{b}$ state
above the $B\bar{B}$ threshold with an established quark model assignment. The $\Upsilon\left(  4S\right)$ is a $1^{--}$ resonance with a
mass
\begin{equation}
m_{b\bar{b}(4S)}=10579.4\pm1.2\text{ MeV} \label{mass4S}%
\end{equation}
and a width%
\begin{equation}
\Gamma_\textup{Expt}\left(  \Upsilon\left(  4S\right)  \right)  =20.5\pm2.5\text{ MeV}
\label{width4S}%
\end{equation}
almost completely determined $\left(  >96\%\right)  $ by its $B\bar{B}$ decay channel.

\begin{figure}
\centering
\includegraphics[width=\linewidth]{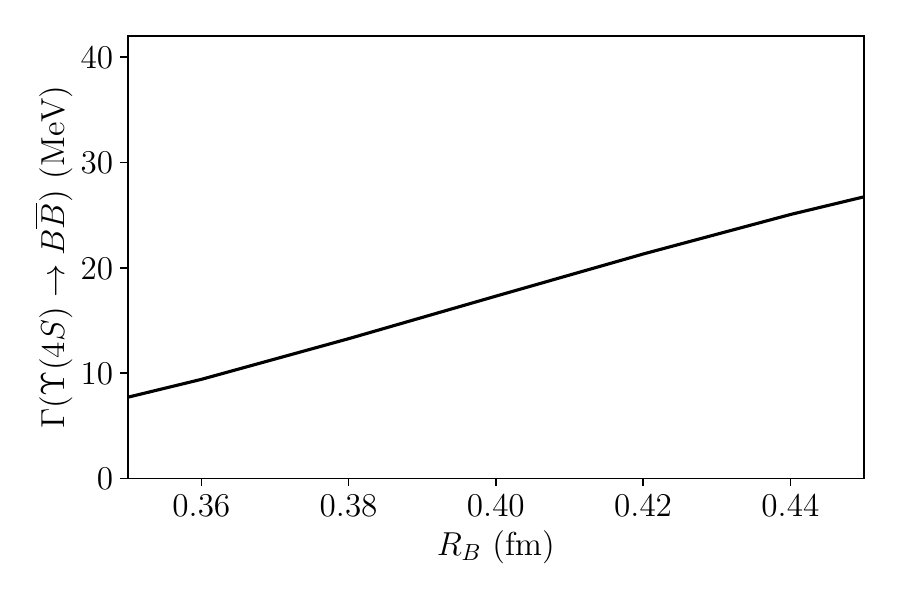}\caption{Calculated
$\Upsilon\left(  4S\right)  \rightarrow B\bar{B}$ width from the static
${^{3}\!}P_{0}$ model.}%
\label{figure4}%
\end{figure}

By neglecting the mass difference between $B^{+}$ and $B^{0}$ we approximate
\begin{equation}
\Gamma_{{^{3}\!}P_{0}}^\textup{st}\left(  \Upsilon\left(  4S\right)  \rightarrow
B^{-}B^{+}\right)  \simeq\Gamma_{{^{3}\!}P_{0}}^\textup{st}\left(  \Upsilon\left(
4S\right)  \rightarrow B^{0}\bar{B}^{0}\right)  \label{BB0widths}%
\end{equation}
so that we can calculate the total decay width%
\begin{align}
&  \Gamma_{{^{3}\!}P_{0}}^\textup{st}\left(  \Upsilon\left(  4S\right)  \rightarrow
B\bar{B}\right)  \label{Totalwidth}\\
&  \equiv\Gamma_{{^{3}\!}P_{0}}^\textup{st}\left(  \Upsilon\left(  4S\right)  \rightarrow
B^{-}B^{+}\right)  +\Gamma_{{^{3}\!}P_{0}}^\textup{st}\left(  \Upsilon\left(  4S\right)
\rightarrow B^{0}\bar{B}^{0}\right)  \nonumber\\
&  \simeq2\Gamma_{{^{3}\!}P_{0}}^\textup{st}\left(  \Upsilon\left(  4S\right)
\rightarrow B^{0}\bar{B}^{0}\right)  \nonumber
\end{align}
by using as the mixing potential $g_{{^{3}\!}P_{0}}(r)$ multiplied by the $\left(
1^{--}\rightarrow0^{-}0^{-}\right)  _{L_{b\bar{b}}=0}$ angular momentum
coefficient. As for the $\Upsilon\left(  4S\right)  $ wave function, we obtain
it by solving the Schr\"{o}dinger equation for $b\bar{b}$ from a standard
Cornell potential \cite{Eic94}
\begin{equation}
V_{C}\left(  r\right)  =\sigma r-\frac{\chi}{r}+m_{b}+m_{\bar{b}}%
-\alpha\label{Cornellpot}%
\end{equation}
with
\begin{align}
\sigma &  =925.6\text{ MeV.fm}^{-1}\label{parpot}\\
\chi &  =102.6\text{ MeV.fm}\nonumber\\
m_{b} &  =5180\text{ MeV}=m_{\bar{b}}.\nonumber
\end{align}
In calculating the relative momentum $k$, we use the Particle Data
Book \cite{PDG} averages for the masses of the decaying $\Upsilon$ meson and $B$ mesons in the final state. The dependence of the calculated width with $R_{B}$ is %
plotted in Fig. \ref{figure4}.

This width is to be compared to the one calculated using $g(r)$ from
(\ref{parg}),
\begin{equation}
\widetilde\Gamma\left(  \Upsilon\left(
4S\right)  \rightarrow B\bar{B}\right)  =19.9\text{ MeV,}
\end{equation}
and to the
experimental one \eqref{width4S}. Notice that
\begin{equation}
\widetilde\Gamma\left(  \Upsilon\left(  4S\right)  \rightarrow B\bar{B}\right)
\simeq\Gamma_\textup{Expt}\left(  \Upsilon\left(  4S\right)  \rightarrow B\bar
{B}\right),  \label{statexp}%
\end{equation}
what can be taken as a proof of the validity of the $m_Q\to\infty$ limit for
$b\bar{b}$ in $\Upsilon\left(  4S\right).$

\begin{figure}
\centering
\includegraphics[width=\linewidth]{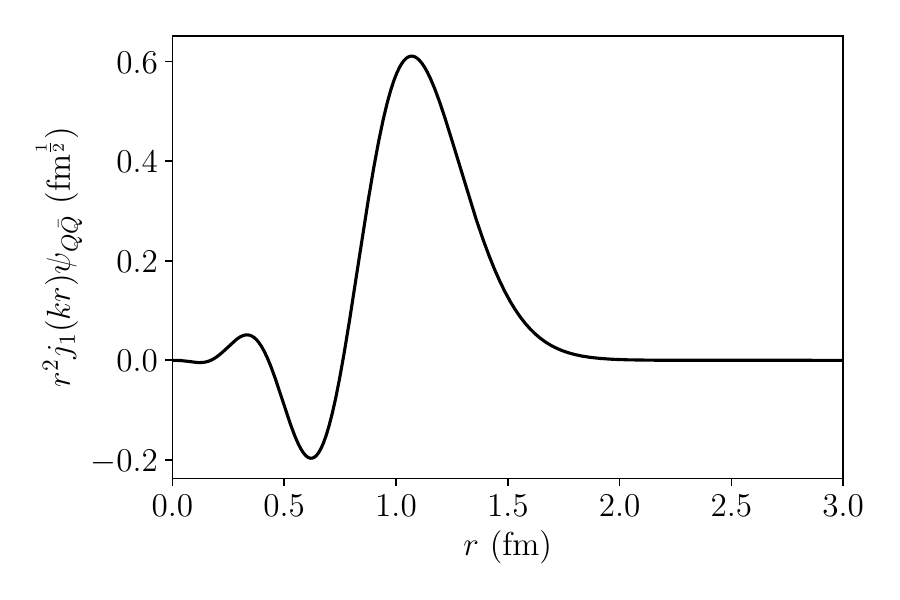}\caption{Radial dependence of
the static ${^{3}\!}P_{0}$ integrand factor multiplying the mixing potential.}%
\label{figure5}%
\end{figure}

As for $\Gamma_{{^{3}\!}P_{0}}^\textup{st}\left(  \Upsilon\left(  4S\right)  \rightarrow
B\bar{B}\right)  $ it is compatible with $\widetilde\Gamma\left(  \Upsilon\left(
4S\right)  \rightarrow B\bar{B}\right)  $ and $\Gamma_\textup{Expt}\left(
\Upsilon\left(  4S\right)  \rightarrow B\bar{B}\right)  $ for an interval
of values of $R_{B}$ around the phenomenological one,%
\begin{align}
& \left[  \Gamma_{{^{3}\!}P_{0}}^\textup{st}\left(  \Upsilon\left(  4S\right)
\rightarrow B\bar{B}\right)  \right]  _{R_{B}\simeq0.40-0.42\text{ fm}%
}\label{3P0exp}\\
& \simeq\Gamma_\textup{Expt}\left(  \Upsilon\left(  4S\right)  \rightarrow
B\bar{B}\right).  \nonumber
\end{align}
We can understand this \textquotedblleft success\textquotedblright\ of the
static ${^{3}\!}P_{0}$ model by examining the integral entering in the calculation
of the width (\ref{Case12}). In Fig. \ref{figure5} we plot the function
multiplying the mixing potential in the integrand. Indeed the short distance behaviour of this function kills the short range
contribution from the mixing potential in the integral what explains the
numerical concordance between $\left[  \Gamma_{{^{3}\!}P_{0}}^\textup{st}\left(
\Upsilon\left(  4S\right)  \rightarrow B\bar{B}\right)  \right]
_{R_{B}\simeq0.40-0.42\text{ fm}}$ and $\Gamma_\textup{Expt}\left(  \Upsilon\left(
4S\right)  \rightarrow B\bar{B}\right)  $. More concretely, the main
contributions to the integral come from $b\bar{b}$ separations close to
the string breaking distance $1.25$ fm.

\begin{figure}
\centering
\includegraphics[width=\linewidth]{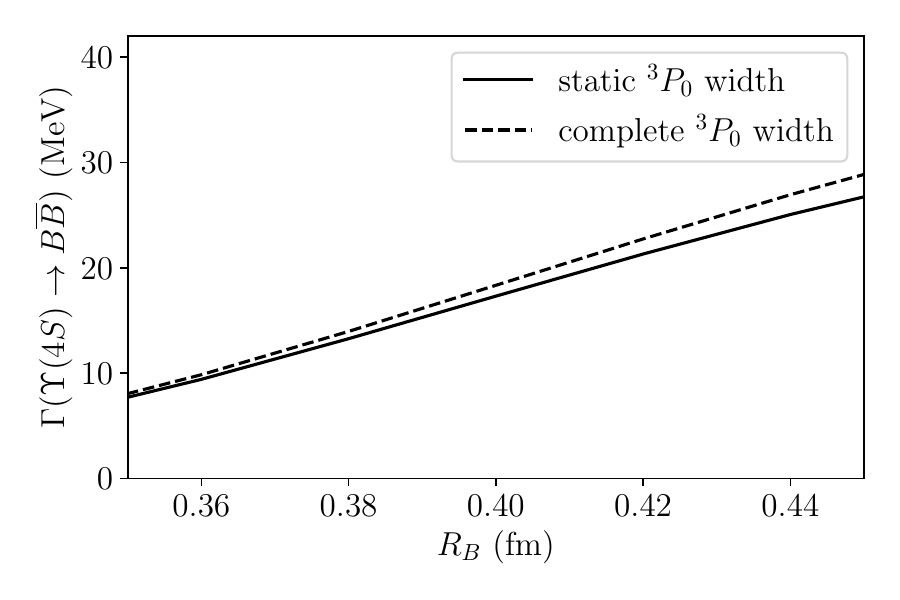}\caption{Static ${^{3}\!}P_{0}$ vs
complete ${^{3}\!}P_{0}$ width for $\Upsilon\left(  4S\right)  \rightarrow
B\bar{B}.$}%
\label{figure6}%
\end{figure}

This \textquotedblleft success\textquotedblright\ of the static ${^{3}\!}P_{0}$
model ($\frac{m_q}{m_Q}=0$) can be translated to the complete ${^{3}\!}P_{0}$ model
($\frac{m_q}{m_Q}>0$) by realizing that $\left[  \Gamma_{{^{3}\!}P_{0}}%
^\textup{st}\left(  \Upsilon\left(  4S\right)  \rightarrow B\bar{B}\right)
\right]  _{R_{B}\simeq0.40-0.42\text{ fm}}$ does not differ significantly from
the complete ${^{3}\!}P_{0}$ width $\left[  \Gamma_{{^{3}\!}P_{0}}\left(
\Upsilon\left(  4S\right)  \rightarrow B\bar{B}\right)  \right]
_{R_{B}\simeq0.40-0.42\text{ fm}}$ calculated from (\ref{I12}) without
neglecting $m_{q}$ against $m_{Q}$. In Fig. \ref{figure6}, we plot both widths
making clear that the ${^{3}\!}P_{0}$ mass correction to the static limit is
pretty small.

Unfortunately the dearth of $b\bar{b}$ data above the $B\bar{B}$
threshold prevents any further accurate checking of the ${^{3}\!}P_{0}$ model in
bottomonium. Apart from $\Upsilon\left(  4S\right)  $ only three other
$b\bar{b}$ states with $1^{--}$ have been observed until now above the
$B\bar{B}$ threshold. Although for one of these extra states,
$\Upsilon\left(  10860\right)  ,$ with a mass $m_{\Upsilon\left(
10860\right)  }=10885.2_{-1.6}^{+2.6}$ MeV the decay to $B\bar{B}$ has
been measured and its leptonic width points out that an $S$-wave component is
present, its mass suggests an intricate mixing of several $S$- and $D$-wave states
through the coupling to different $B^{\left(  \ast\right)  }\bar
{B}^{\left(  \ast\right)  }$ thresholds whose study is out of the scope of this article.

\section{Analysis of the ${^{3}\!}P_{0}$ model in charmonium}

\begin{figure}[ptb]
\centering
\includegraphics[width=\linewidth]{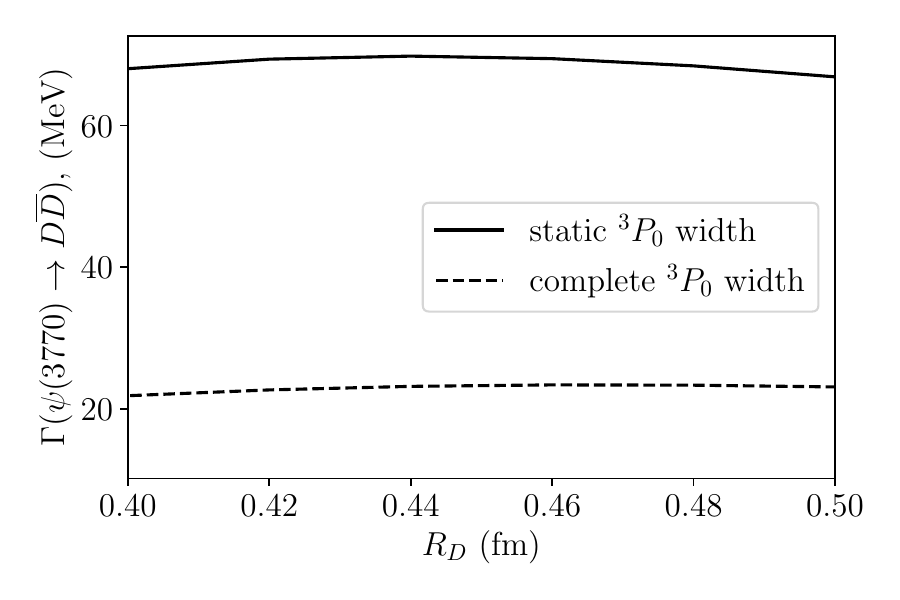}\caption{Static ${^{3}\!}P_{0}$ vs
complete ${^{3}\!}P_{0}$ width for $\psi\left(  3770\right)  \rightarrow
D\bar{D}.$}%
\label{figure7}%
\end{figure}

The finite-mass corrections in the complete ${^{3}\!}P_{0}$ model may be much more important for charmonium than bottomonium, given that the charm quark mass $m_c$ is roughly three times smaller than the bottom quark mass $m_b$. When
trying to check this we face the difficulty of not having any $c\bar{c}$
state above the $D\bar{D}$ threshold with an established quark model
assignment. To circumvent this absence we shall assume, in line with the
quark model mass prediction, that the lower resonance above threshold, called
$\psi\left(  3770\right)  $ in \cite{PDG}, is the $1D$ state
of $c\bar{c}$. The $\psi\left(  3770\right)  $ is  a $1^{--}$ resonance with a mass%
\begin{equation}
m_{\psi\left(  3770\right)  }=3773.7\pm0.7\text{ MeV} \label{mpsi}%
\end{equation}
and a width%
\begin{equation}
\Gamma_\textup{Expt}\left(  \psi\left(  3770\right)  \right)  =27.2\pm1.0\text{ MeV}
\label{widthpsi}%
\end{equation}
almost completely determined $\left(  93_{-9}^{+8}\text{ }\%\right)  $ by its $D\bar{D}$ decay channel.\footnote{In this regard, we should mention that the measured
leptonic decay width of $\psi\left(  3770\right)  $ indicates a small component of the $2S$ state.} Then, if we
neglect the mass difference between $D^{\pm}$ and $D^{0}$ or $\bar{D}%
^{0}$, we can use the string breaking transition rate $g(r)$ from (\ref{parg})
multiplied by the corresponding $\left(  1^{--}\rightarrow0^{-}0^{-}\right)
_{L_{c\bar{c}}=2}$ angular momentum coefficient to calculate the width. We
obtain
\begin{equation}
\widetilde\Gamma\left(  \psi\left(  3770\right)  \rightarrow D\bar{D}\right)
=88.1\text{ MeV}.%
\end{equation}
This value, more than three times bigger than the experimental one indicates
the failure of the limit $m_Q\to\infty$ for charmonium and the need for mass
corrections. Indeed, from the static ${^{3}\!}P_{0}$ model an indirect charm quark
mass correction can be introduced in the transition rate through the $D$ or
$\bar{D}$ harmonic wave functions with $R_{D}>R_{B}$ as expected from phenomenology:%

\begin{equation}
\left\vert g_{{^{3}\!}P_{0}}^{c\bar{c}}(r)\right\vert =\gamma\frac{\sqrt{\pi}}{8}%
\frac{r}{R_{D}^{2}}e^{-\frac{r^{2}}{4R_{D}^{2}}}%
\end{equation}
with $\gamma=2.24$ and $R_{D}=0.4-0.5$ fm. Thus, the calculated value of the width
from $g_{{^{3}\!}P_{0}}^{c\bar{c}}(r)$ that we still shall call static ${^{3}\!}P_{0}$ width is
significantly reduced as compared to $\widetilde\Gamma\left(  \psi\left(  3770\right)  \rightarrow D\bar{D}\right)$ as
shown in Fig. \ref{figure7}.

Furthermore, if we implement further mass corrections through the use of the
complete ${^{3}\!}P_{0}$ model ($\frac{m_q}{m_Q} > 0$) the calculated width from
(\ref{I12}) becomes closer to the experimental one as also shown in Fig.
\ref{figure7}.

Therefore, although no other data is available in charmonium for further
checking we may tentatively conclude that the ${^{3}\!}P_{0}$ model allows for the
implementation of the dominant corrections to the $m_Q\to\infty$ limit.

\section{Summary and conclusions}

The phenomenological success of the
${^{3}\!}P_{0}$ model for strong decays of hadrons into pairs of hadrons is especially puzzling given its tenuous connection with QCD. In this paper, we have provided a theoretical underpin to the ${^{3}\!}P_{0}$ model for decays of
quarkonia into open flavor two-meson states. In the heavy-quark limit,
this description can be compared to the one in terms of
mixing potentials between quarkonium and heavy-meson pairs accessible from lattice QCD studies
of string breaking. We have shown, by fitting the value of the ${^{3}\!}P_{0}$
constant pair poduction amplitude and the harmonic oscillator parameter
describing the final meson state, that the transition rates for both
descriptions are equivalent for quark-antiquark separations around and
beyond the string breaking distance. Inasmuch as these separations dominate
the decay of quarkonium into heavy-meson pairs, this provides a QCD based justification for
the use of the ${^{3}\!}P_{0}$ model with a constant pair creation amplitude. We have also shown that a fit to the full range of heavy quark-antiquark separations may indicate a generally non-constant pair creation amplitude. In practice, we have
explicitly shown these results in bottomonium by studying the $\Upsilon\left(
4S\right)  \rightarrow B\bar{B}$ decay. Regarding charmonium
decays, finite-mass corrections beyond the heavy-quark limit play a more prominent role. We
have explicitly shown, through the analysis of the decay $\psi\left(  3770\right)
\rightarrow D\bar{D}$, that the finite-mass corrections implicitly
incorporated in the ${^{3}\!}P_{0}$ model improve the quality of the description significantly. Using the results of Ref.~\cite{Braa24}, this analysis could be extended to decays of quarkonia into some final states containing excited heavy mesons with positive parity. From these results, we
conclude that the ${^{3}\!}P_{0}$ model grasps the essential physics of the
string breaking mechanism in QCD underlying the decays of quarkonia into pairs of heavy mesons.

\begin{acknowledgments}
This work has been supported by Conselleria de Innovaci\'{o}n, Universidades,
Ciencia y Sociedad Digital, Generalitat Valenciana GVA PROMETEO/2021/083,
by EU Horizon 2020 Grant No. 824093 (STRONG-2020), and by MCIU/AEI Severo Ochoa project CEX2023-001292-S. T.T. acknowledges the
support of CONICET of Argentina. R.B. acknowledges the support of the U.S. Department of Energy under grant DE-SC0011726.
\end{acknowledgments}


\begin{thebibliography}{99}                                                                                               %


\bibitem {Mic69}L. Micu, Nucl. Phys. B \textbf{10}, 521 (1969).

\bibitem {LY}A. Le Yaouanc, L. Oliver, O. P\`{e}ne, and J.-C. Raynal, Phys.
Rev. D \textbf{8}, 2223 (1973); Phys. Lett. B71, 397 (1977); Phys. Lett. B72,
57 (1977).

\bibitem {Ono}S. Ono, Phys. Rev. D \textbf{23}, 1118 (1981).

\bibitem {BGS}T. Barnes, S. Godfrey, and E.S. Swanson, Phys.Rev.D 72, 054026 (2005).

\bibitem {Bal05}G. S. Bali, H. Neff, T. D\"{u}ssel, T. Lippert, and K.
Schilling, Phys. Rev. D 71, 114513 (2005).

\bibitem {Bul19}J. Bulava, B. H\"{o}rz, F. Knechtli, V. Koch, G. Moir, C.
Morningstar, and M. Peardon,\ Phys. Lett. B 793, 493 (2019).

\bibitem {Bic20}P. Bicudo, M. Cardoso, N. Cardoso, and M. Wagner, Phys.Rev.D
101, 034503 (2020).

\bibitem {Fan61}U. Fano, Phys. Rev. 124, 1866 (1961).

\bibitem {Eic78}E. Eichten, K. Gottfried, T. Kinoshita, K. D. Lane, and T. M.
Yan, Phys. Rev. D 17, 3090 (1978), [Erratum: Phys. Rev. D21, 313 (1980)]

\bibitem {Bru21a}R. Bruschini and P. Gonz\'{a}lez, Phys.Rev.D 103, 074009 (2021).

\bibitem {Bru20}R. Bruschini, P. Gonz\'{a}lez, Phys.Rev.D 102, 074002 (2020).

\bibitem {Bru23}R. Bruschini, JHEP 08, 219 (2023).

\bibitem{Braa24} E. Braaten and R. Bruschini, Phys.Rev.D 109, 094051 (2024).

\bibitem {Bru21b}R. Bruschini and P. Gonz\'{a}lez, Phys.Rev.D 103, 114016 (2021).

\bibitem {PDG}S. Navas et al. (Particle Data Group), Phys. Rev. D 110, 030001 (2024).

\bibitem {Eic94}E. Eichten and C. Quigg, Phys.Rev.D 49 , 5845 (1994).


\end{thebibliography}
\end{document}